\documentclass[aps,pre,preprint,showpacs,amssymb]{revtex4}
\usepackage{amsmath,bm,amssymb,graphics,epsfig}
\usepackage{amsmath,graphics,epsfig}

\newcommand{\gae}{\hbox{\lower0.7ex\hbox{$\sim$}\llap{\raise0.4ex\hbox{$>$}}}}
\newcommand{\lae}{\hbox{\lower0.7ex\hbox{$\sim$}\llap{\raise0.4ex\hbox{$<$}}}}
\newcommand{\half}{\frac{1}{2}}
\begin{document}
\title{Equivalent-neighbor Potts models in two dimensions}
\author{Xiaofeng Qian~$^1$, Youjin Deng~$^{2,4}$~\footnote{Corresponding
author; email: yjdeng@ustc.edu.cn},
Yuhai Liu~$^{3}$, Wenan Guo~$^{3,5}$, and Henk W.~J. Bl\"ote~$^{1}$ }
\affiliation{$^1$Lorentz Institute, Leiden University,
  P.O. Box 9506, 2300 RA Leiden, The Netherlands}
\affiliation{$^{2}$ Hefei National Laboratory for Physical Sciences at
Microscale, Department of Modern Physics, University of Science and
Technology of China, Hefei 230027, China }
\affiliation{$^{3}$Physics Department, Beijing Normal University,
Beijing 100875, China}
\affiliation{$^{4}$State Key Laboratory of Theoretical Physics,
Institute of Theoretical Physics, Chinese Academy of Sciences,
Beijing 100190, China}
\affiliation{$^{5}$  Beijing Computational Science Research Center,
Beijing 100084, China.}
\date{\today}
 \begin{abstract}
We investigate the two-dimensional $q=3$ 
and 4 Potts models with a variable interaction range by means of Monte
Carlo simulations. We locate the phase transitions for several interaction
ranges as expressed by the number $z$ of equivalent neighbors. For not too 
large $z$, the transitions fit well in the universality classes of the 
short-range Potts models. However, at longer ranges the transitions become 
discontinuous. For $q=3$ we locate a tricritical point separating the 
continuous and discontinuous transitions near $z=80$, and a critical fixed
point between $z=8$ and 12. For $q=4$ the transition becomes discontinuous for
$z > 16$. The scaling behavior of the $q=4$ model with $z=16$ approximates
that of the $q=4$ merged critical-tricritical fixed point predicted by the
renormalization scenario.
 \end{abstract}
\maketitle
 
\section {Introduction}
\label{sec1}
In phase transitions, the range of the interactions plays an important role. 
Models with interactions decaying as a negative power $-p$ of the distance 
appear to display a considerable variety of universality classes as a function 
of $p$, and as a function of the dimensionality \cite {FMN,Sak,LB1}. For large 
$p$ the interactions decay fast and one finds the usual short-range universal 
behavior. For sufficiently small $p$ the interactions decay only slowly and
one finds mean-field-like critical behavior. For intermediate values of
$p$ the critical exponents may depend continuously on $p$.

\begin{figure}[bthp]
\begin{center}
\leavevmode
\epsfxsize 8.4cm
\epsfbox{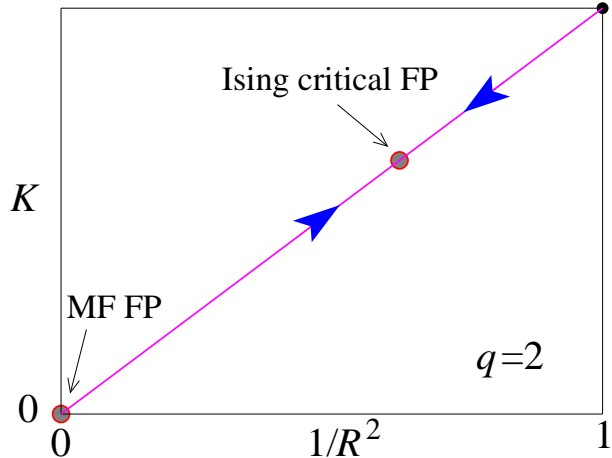}
\end{center}
\caption{Schematic renormalization flow diagram along the line of phase 
transitions of the $q=2$ Potts model with medium-range interactions.
The critical line connecting the mean-field (MF) fixed point and the
Ising fixed point is parametrized by the range $R$ of
the interactions. The finiteness of the interaction range is relevant
at the MF fixed point and leads to crossover to the Ising fixed point.
}
\label{ising}
\end{figure}

A different way to modify the range of the interactions is specified in the
so-called equivalent-neighbor models, in which the pair interactions are
constant up to a range $R$ and then abruptly fall to zero.
Following Ref.~\onlinecite{LBB}, we refer
to these models as medium-range models.  In the limit $R \to \infty$,
the equivalent-neighbor model reduces to the mean field model;
for sufficiently small $R$ it will naturally display the usual short-range 
universal behavior. But it seems that the analogy with the case 
of power-law decay of interactions ends here. Medium-range Ising models, 
with interactions of a variable range $R$ display uniform crossover from
the vicinity of the mean-field fixed point at $R=\infty$ to the Ising 
critical fixed point at small $R$ \cite{LBB}. The model belongs to the Ising
universality class for all finite $R$. The scaling field parametrizing
the crossover phenomenon is the irrelevant Ising temperature field.
The flow diagram for the Ising model is shown in Fig.~\ref{ising}.

The question naturally arises whether such a uniform crossover also
occurs in the more general context of the $q$-state Potts model \cite{P},
of which the Ising model is the special case with $q=2$. Another
possibility is suggested by the renormalization scenario for the
two-dimensional dilute $q$-state Potts model \cite{NBRS} with $0<q<4$.
In this context the leading irrelevant field, parametrizing the
critical surface in parameter space, is controlled by the activity
of the vacancies. When the latter parameter is increased, while
adjusting the Potts coupling to maintain criticality, a threshold
occurs where the model becomes tricritical. Beyond the threshold the
ordering transition becomes discontinuous. If the parameters controlling 
the irrelevant fields  of the dilute and the equivalent-neighbor Potts
models are  sufficiently analogous for $q>2$, then this scenario, i.e., a
tricritical point separating a range of critical and a range of first-order
transitions, might also occur for the equivalent-neighbor Potts model. 
This is not a new idea. It was already raised by
Hilhorst \cite{HJH} in a discussion following the renormalization
results for the Potts model with vacancies \cite{NBRS}.

This possibility is also in line with work of Biskup et al.~\cite{Bi} which
concerns $q=3$ models with interactions whose strength decays smoothly
to zero at infinite range. For a sufficiently slow decay, a first-order
transition is predicted. It is also in line with results of Gobron and Merola
\cite{GM} for the  mean-field Potts model perturbed with a Kac potential.
In order to provide quantitative answers to the similar question for a simple
Potts model system, we consider the equivalent-neighbor model with a finite
but variable interaction range $R$, described by the reduced Hamiltonian
\begin{equation}
{\cal H}/k_{\rm B}T = -K \sum_{i<j } \theta(R-r_{ij})\delta_{\sigma_i,
\sigma_j} \hspace{5mm} (\sigma_i=1,\cdots,q) \, ,
\label{model}
\end{equation}
where the Potts variables $\sigma_i$ carry indices that refer to the sites 
of a square lattice with periodic boundary conditions.
Interacting pairs of sites are selected by the step function $\theta$
(defined by $\theta(x)=1$ for $x\geq 0$ and $\theta(x)=0$ for $x< 0$).
Thus, interactions of strength $K$ occur with all neighbors within a
distance $R$. 
In this work, we specify the interaction range $R \propto z^{1/d}$ by
the number $z$ of equivalent-neighbors interacting with a spin on a
$d$-dimensional lattice.

In particular we investigate the $q=3$ and the $q=4$ Potts model on
$L \times L$ square lattices for a sequence of finite sizes $L$.
This task is performed numerically, by means of a cluster Monte Carlo
method \cite{LB} that is especially suitable for this problem,
because it not only reduces critical slowing down, but it also remains
very efficient for systems with interactions of a long range.
During the simulations, we sampled the densities $\rho_i$ of the Potts
variables in state $i$, from which we obtained the magnetization moments 
and the Binder ratio \cite {KB}, as explained in Sec.~\ref{sec2}.
We use finite-size scaling  \cite{FSS} to analyze these data to obtain
the location of the phase transitions and their universality classes.
In Sec.~\ref{sec3} we show the results for the $q=3$ Potts model, for
several values of $z$ in the range $4<z<120$. Results for several $q=4$
Potts models with $4<z<60$ are presented in Sec.~\ref{sec4}. Finally, 
discussions and conclusions are listed in Sec.~\ref{sec5}.
The main results of the present article appeared earlier in the 
PhD thesis of one of us \cite{xfth}.

\section {Methods and sampled quantities}
\label{sec2}
The principle of the Monte Carlo technique employed for the study of
the present two-dimensional medium-range Potts models was explained in
detail in Ref.~\onlinecite{LB} for the Ising case $q=2$, and it can be
trivially generalized to $q>2$ Potts models. The algorithm is organized
such that it requires a computer time that is almost independent
of the number $z$ of interacting neighbors per spin. We used the
Wolff-like single-cluster version \cite{Wolff} of the algorithm.

Since the locations of the phase transitions are unknown for general $z$,
our first task is to determine them. This determination is based on
the Monte Carlo sampling of the moments of the magnetization
density $m$. This quantity depends on the densities $\rho_i$ of the
Potts states $i=1,2,\cdots ,q$ as
\begin{equation}
m^2\equiv\frac{1}{q-1}\sum_{i=1}^{q-1}\sum_{j=i+1}^q(\rho_i-\rho_j)^2 \, .
\label{m2def}
\end{equation}
This definition is in accordance with the interpretation of the Potts
spins as vectors with $q$ equivalent orientations
in $q-1$ dimensional spin space.
The magnetization moments determine a dimensionless ratio $Q$, 
related to the Binder cumulant \cite{KB}, defined as:
\begin{equation}
 Q=\frac{\langle m^2\rangle ^2}{\langle m^4\rangle}\, .
\end{equation}
The expected finite-size scaling behavior of $Q$ near the transition
point is obtained by expansion of the scaling formula for the free energy
near the pertinent critical or tricritical fixed point. This leads to
\begin{displaymath}
Q(K,L)= Q_0 + \sum_k a_k(K-K_{\rm c})^k L^{k y_t} + \sum_j b_j L^{y_j} 
\end{displaymath}
\begin{equation}
+ c (K-K_{\rm c}) L^{y_t+y_1}+d(K-K_{\rm c})^2 L^{y_t}+\cdots \, ,
\label{drq}
\end{equation}
where $Q_0$ is a universal constant, $y_t$ is the renormalization exponent
describing the scaling of the temperature field, and the $y_j$ with
$j=1,2,\cdots$ are
negative exponents describing corrections that will be discussed later,
and the $a_k$, the $b_j$, $c$, and $d$ are unknown amplitudes. The term
with amplitude $d$ describes the nonlinearity of the temperature field as
a function of $K$.

In the case of the four-state Potts model, the behavior is less simple
because of the presence of a marginal operator, as predicted
by the renormalization scenario due to Nienhuis et al.~\cite{NBRS}.
From a further analysis of the renormalization equations \cite{NS,CNS,SS},
it is possible to predict the finite-size-scaling
behavior of the singular part of free energy  as a function of the
temperature scaling field $t \simeq K-K_{\rm c}$, the magnetic scaling field
$h$, and the marginal field $v$ as
\begin{equation}
f_s(t,h,v,L^{-1}) =L^{-d} f_s(L^{y_t}u^{3/4} t,L^{y_h}u^{1/16} h, uv,1) \, ,
\label{fq4}
\end{equation}
where $u(L) \equiv 1/[1-(v/\pi) \ln L]$.
Since the magnetization moments can be expressed in terms of derivatives
of the free energy with respect to the magnetic field, one can also obtain
the expected scaling behavior of $Q$. In leading orders one finds that,
for $K=K_{\rm c}$,
\begin{equation}
Q(K_{\rm c},L)=Q_0+ c_1/(1-b\ln L)+c_2/(1- b\ln L)^2+c_3/(1- b\ln L)^3+
\sum_j b_j L^{y_j} + \ldots
\label{Qq4c}
\end{equation}
where $b \propto v$, and $c_k \propto v^k$, thus $b \propto c_1$ as well.
The proportionality constants are universal but unknown.
The finite-size scaling behavior of $Q$ near the transition point
follows by additional differentiation of $f_s$ to the temperature field as
\begin{equation}
Q(K,L)= Q(K_{\rm c},L) + \sum_k q_k(K-K_{\rm c})^ku^{3k/4} L^{k y_t} +
\sum_j b_j L^{y_j} +\ldots
\label{drq4}
\end{equation}
where $q_k \propto v^{3k/4}$, with universal but unknown proportionality
constants. 

The ratio $Q$ is a useful quantity to locate phase transitions and to
determine the associated temperature-like exponent. From Eq.~(\ref{drq})
one finds that the $Q$ versus $K$ curves for different values of the
finite-size parameter $L$ intersect at values approaching
$K=K_{\rm c}$ for large $L$. Moreover, the slopes of these curves are
asymptotically proportional to $L^{y_t}$, which thus allows the
estimation of $y_t$.

\begin{figure}
\begin{center}
\leavevmode
\epsfxsize 12.0cm
\epsfbox{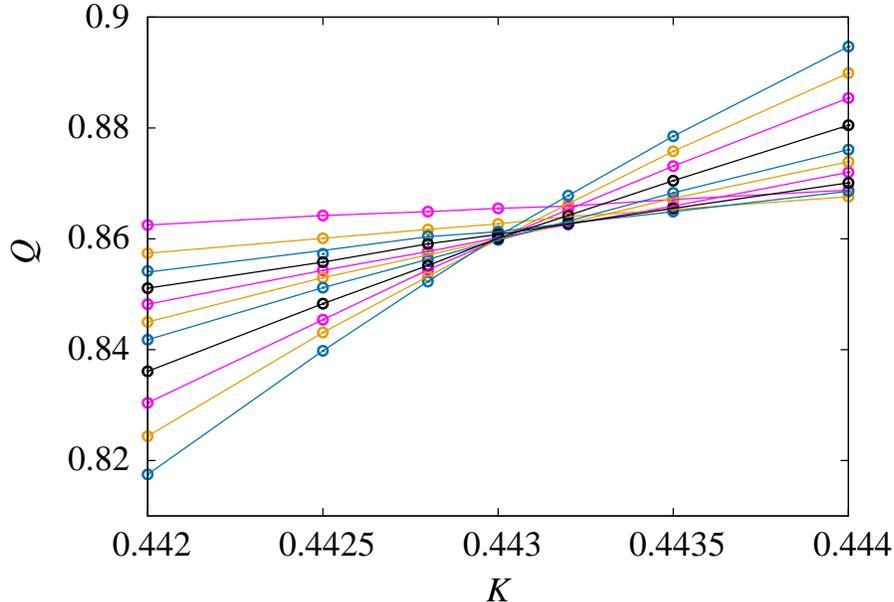}
\end{center}
\caption{The Binder ratio $Q$ of the three-state Potts model with
8 equivalent interacting neighbors vs. coupling $K$, for system sizes
$L=6$, 9, 12, 15, 18, 21, 24, 30, 36, 42, and 48. Larger system sizes
correspond with steeper curves.}
\label{k8n}
\end{figure}

For each model, simulations were performed for several system sizes 
in a suitable range of $K$ near criticality, and $6\times 10^6$  or
more samples
were taken for each data point specified by $q$, $K$ and $z$. The
intersections of finite-size data for $Q$ versus $K$, taken at 
different values of $L$ but for the same $q$ and $z$, reveal the 
location of the critical point. This is illustrated in Fig.~\ref{k8n} 
for the $q=3$ Potts model with $z=8$, i.e., nearest- and next-nearest
neighbor interactions. 
A more accurate location was determined by a least-squares 
analysis according to Eq.~(\ref{drq}). Similar analyses were performed
for the other choices of $q$ and $z$ investigated in the present work. 

We have also searched for possible evidence, in the form of hysteresis
loops, for first-order transitions at a finite interaction range. We 
thus determined the behavior of the energy and the magnetization while
the coupling $K$ was changed in small steps separated by long intervals.
Furthermore we investigated the autocorrelation time $\tau$, and 
the distributions $p(E)$ of the energy and $p(m)$ of the magnetization. 

\section {Results for three-state Potts models}
\label{sec3}
\subsection {Location and nature of the phase transitions}
The Binder ratio $Q$ is assumed to be universal for critical Potts
models with the same $q$, but this universal number still depends on the
geometry of the finite system. The relevant factors are the ratio of the
microscopic couplings in different directions, the boundary conditions
and the shape of the system, for instance the aspect ratio of a
rectangular periodic box.
In this work we restrict ourselves to systems with square symmetry, which
pertains to the lattice, the couplings and to the boundary conditions.
The universal value of $Q_0$ can therefore conveniently be determined 
from the nearest-neighbor Potts model, for which we know the exact
critical point as $K_{\rm c}=\ln(1+\sqrt{3})$. We therefore simulated the
nearest-neighbor three-state Potts model at the critical point, using
square systems with sizes $L=6$, 7, 8, ..., 280, 320. We fitted the
finite-size data by
Eq.~(\ref{drq}), using the known values of the critical point and the
critical exponents \cite{BN}, of which the temperature exponent is
$y_t=6/5$ and the leading irrelevant exponent $y_1=-4/5$.
This leads to $Q_0=0.85410\,(10)$.

As a consistency check, we also simulated the dilute Potts model to
determine $Q_0$ for  the three-state Potts model
near the critical fixed point, which is located \cite{QDB} 
at $K_{\rm cfp}=1.16941(2)$, $D_{\rm cfp}=1.376483(5)$.
At the critical fixed point, the leading correction term with exponent
$-4/5$ is suppressed.
For the model at the critical fixed point, we simulated 21 systems
$L=6$, 8, 10, ..., 150, 210, and obtain $Q_0=0.85408\,(7)$ and
$y_1=-1.13\,(4)$. This value is close to an expected correction exponent
$X_{h1}-X_{h2}=-6/5$. The values of the magnetic exponents $X_{h1}$ and
$X_{h2}$ are given in Ref.~\cite{BN}. When we fix the correction exponent
at the value $y_1=-6/5$, we obtain $Q_0=0.85412(5)$.
These relatively accurate results for $Q_0$  will be useful for the
analysis of models with more neighbors.

We also simulated the dilute Potts model to
determine $Q_0$ at the tricritical point of the
three-state Potts model, which is located \cite{QDB}
near $K_{\rm tfp}=1.649903$, $D_{\rm tfp}=3.152152$.
For the tricritical dilute Potts model we used system sizes $L=6$,
8, 10, ..., 84, 108.  Fits with fixed $y_t=12/7$ \cite{BN}
and $y_1$ left as a free parameter show that there exists a correction
to scaling with an exponent near $y_1=-1$, with an uncertainty  margin of a
few tenths. This exponent is consistent with $y_1=X_{h1}-X_{h2}=-6/7$.
The $\chi^2$ criterion provides strong indications that another correction to
scaling is present, but the data are not accurate enough to allow a reliable
determination of its exponent. Further
corrections may be expected with exponent $y_2=-10/7$
(irrelevant exponent $X_{14}$ in the Kac table) and with $d-2y_h=-38/21$.
The resulting values for the $Q_0$ are still somewhat dependent on which
correction exponent is included.  Taking into account the uncertainty due
to this dependence, as well as the statistical error margin,
the estimated result is $Q_0=0.743\,(4)$.

For several values of $z$, we determine the critical points, and also 
estimate the temperature exponent by least square fits. The results are
included in Table \ref{8pot3}.
\begin{table}[htbp]
\centering
\caption{\label{8pot3}
Results for the Binder ratio $Q_0$ and thermal exponent $y_t$ for $q=3$
models for several ranges of interaction. These results are obtained by fits
of Eq.~(\ref{drq}) to the Monte Carlo data, with all parameters left free,
except $K_{\rm c}$ in the case of the nearest-neighbor model.
The tricritical point lies in the neighborhood of $z=80$, because the
result for $y_t$ is closest to the tricritical value $y_t=12/7$ for $q=3$.
For smaller $z$ the results tend to the critical value $y_t=6/5$,
and for larger $z$ to the discontinuity fixed point value $y_t=2$ which
applies to first-order transitions.
The third column ``Ms'' lists the number of millions of samples taken
per data point.
The error estimates between parentheses are based on two standard
deviations in the statistical analysis.}
\vskip 5mm
\begin{tabular}{|c|c|c|c|l|l|l|l|}
\hline
 $z$&  $L$  &Ms &  $K_{\rm c}$     &   $Q_0$     &  $y_t$    & $y_1$   &$y_2$\\
\hline
  4 & 6-320 &25 & $\ln(1+\sqrt{3})$(exact)&0.8542 (1)&1.20 (3)&$-$4/5 &$-$6/5\\
  8 & 6-240 & 8 & 0.442907  (3)   & 0.8536 (8) & 1.18  (2) &$-$4/5    &$-$6/5\\ 
 12 & 6-240 & 6 & 0.272027  (2)   & 0.8537 (4) & 1.197 (6) &$-$4/5    &$-$6/5\\
 20 & 6-240 & 8 & 0.154075  (2)   & 0.852  (4) & 1.19  (2) &$-$4/5    &$-$6/5\\
 28 & 9-240 & 8 & 0.106430  (2)   & 0.848  (4) & 1.15  (3) &$-$4/5    &$-$6/5\\
 36 & 9-270 & 8 & 0.081432  (2)   & 0.853  (6) & 1.18  (3) &$-$4/5    &$-$6/5\\
 48 & 9-270 & 8 & 0.060112  (2)   & 0.838 (16) & 1.24  (4) &$-$4/5    &$-$6/5\\
 56 & 9-360 &10 & 0.051188  (2)   & 0.802  (6) & 1.36  (4) &$-$4/5    &$-$6/5\\
 68 &12-600 &11 & 0.0418853 (8)   & 0.773  (4) & 1.45  (4) &$-$4/5    &$-$6/5\\
 80 &12-600 & 8 & 0.0354315 (4)   & 0.753  (2) & 1.64  (4) &$-$4/5    &$-$2  \\
 100&18-160 & 6 & 0.0282084 (4)   & 0.744  (8) & 1.98  (6) &$-$1      &$-$2  \\
 120&18-120 & 6 & 0.0234324 (4)   & 0.754  (8) & 2.01  (5) &$-$1      &$-$2  \\
\hline
\end{tabular}
\end{table}
The dependence of the estimates of $y_t$ and of $Q_0$ for different $z$
provides some information  on the nature of the phase transition.
For $z \leq 48$, the results are consistent with
the universality class of the $q=3$ short-range model. It is however
clear that crossover phenomena occur near $z=80$, affecting the accuracy 
of the results and their error estimation. In particular the results for
$y_t$ and $Q_0$ near $z=56$ display this effect.
The results for $z=80$ lie 
close to the tricritical values given above.  For $z>80$ the
results are consistent with first-order behavior: the value $y_t=2$ 
corresponds with the discontinuity fixed point exponent \cite{NN},
and the universal ratio is expected to satisfy $Q_0=3/4$ at the
coexistence of three
ordered phases and one disordered phase \cite{Yuhai}.
More accurate estimations of critical points were obtained when the Binder
ratio and the temperature exponents were fixed at their expected values.
The results are listed in Table \ref{pot32}.
\begin{table}[htbp]
\centering
\caption{\label{pot32}
Transition points $K_{\rm c}$ for three-state Potts models as determined by
least-squares fits with $y_t$ fixed at $6/5$ for $z<80$, and at $y_t=2$
for $z>80$.  For $z=80$, $y_t$ was fixed at $12/7$ although the data in 
Table \ref{8pot3} suggests that the tricritical value of $z$ may be
slightly larger than 80.
We fixed $Q_0=0.85412$ for $z < 80$,  $Q_0=0.743$ for $z=80$, and $Q_0=0.75$
for $z \ge 100$. For $z=4$, we used the exact critical value of $K_{\rm c}$.
The error margins are based on two standard deviations in the statistical
analysis.}
\vskip 5mm
\begin{tabular}{|c|c|c|c|c|l|l|c|}
\hline
 $z$ &$L_{min}$&$K_{\rm c}  $    & $Q_0$&$y_t$& $y_1$  &$y_2$& $b_1$  \\
\hline
  4  &  6 &$\ln(1+\sqrt{3})$(exact)&0.85412&6/5&$-$4/5&$-6/5$&   0.148 (2) \\
  8  &  6 & 0.4429080 (10) & 0.85412 &6/5  &$-$4/5    &$-6/5$&$ $0.085 (5) \\
 12  &  6 & 0.2720275 (6)  & 0.85412 &6/5  &$-$4/5    &$-6/5$&$-$0.155 (2) \\
 20  &  9 & 0.1540760 (5)  & 0.85412 &6/5  &$-$4/5    &$-6/5$&$-$0.68  (2) \\
 28  &  9 & 0.1064309 (4)  & 0.85412 &6/5  &$-$4/5    &$-6/5$&$-$1.94  (5) \\
 36  &  9 & 0.0814320 (4)  & 0.85412 &6/5  &$-$4/5    &$-6/5$&$-$3.37  (5) \\
 48  &  9 & 0.0601132 (3)  & 0.85412 &6/5  &$-$4/5    &$-6/5$&$-$7.85  (7) \\
 56  & 12 & 0.0511894 (2)  & 0.85412 &6/5  &$-$4/5    &$-6/5$&$-$14.3  (8) \\
 68  & 60 & 0.0418858 (2)  & 0.85412 &6/5  &$-$4/5    &$-6/5$&$-$40    (6) \\
 80  & 48 & 0.03543150(6)  & 0.743   &12/7 &$-$4/5    &$-2  $&$ $1.5   (4) \\
 100 & 18 & 0.0282086 (1)  & 3/4     &2    &$-$1      &$-2  $&$-$2.3   (5) \\
 120 & 18 & 0.0234323 (1)  & 3/4     &2    &$-$1      &$-2  $&$-$2.4   (4) \\
\hline
\end{tabular}
\end{table}

\begin{figure}
\begin{center}
\leavevmode
\epsfxsize 15.0cm
\epsfbox{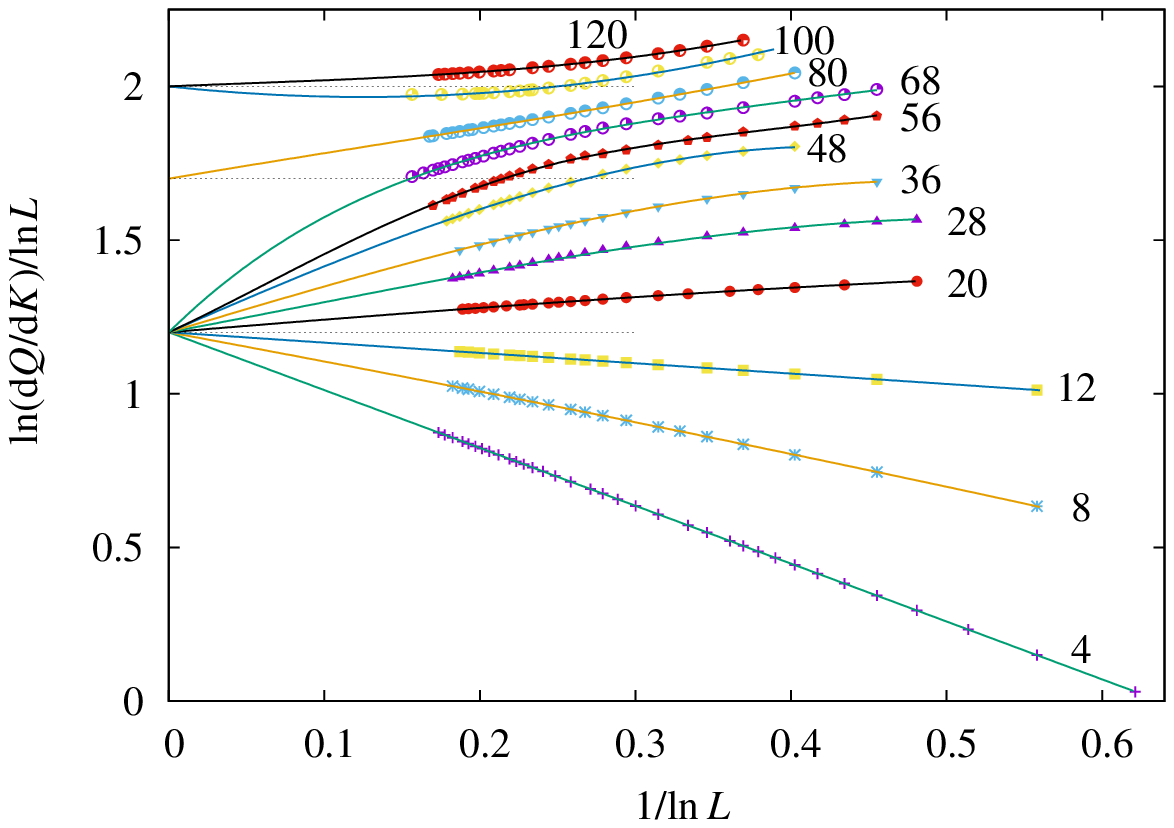}
\end{center}
\caption{Finite-size dependence of the derivative of the Binder ratio $Q$
of the three-state Potts model with respect to $K$, taken at $K_{\rm c}$.
The quantity plotted along the vertical scale is defined in the text
and chosen such that the data should converge, for sufficiently large $L$,
to the temperature exponent $y_t$ which is 6/5 for the three-state critical 
Potts model, 12/7 for the tricritical three-state Potts model, and 2 for
the first-order range. These values are shown by thin horizontal lines.
The data points for each value of $z$ are connected by a curve which is
also intended to guide the eye to the limiting value at $L=\infty$ on the
vertical scale, according to our interpretation of the data.
The value of $z$ is indicated in the figure for each curve. These results
show that the model with $z=80$ lies close to the tricritical point.
}
\label{q3rg}
\end{figure}

In order to shed more light on the crossover phenomenon near $z=80$, 
we study the first derivative of $Q$ with respect to the 
coupling $K$ at criticality, which can be derived from Eq.~(\ref{drq}) as 
\begin{equation}
\label{dQ}
\left. {\frac{{\rm d}Q}{{\rm d}K}}\right|_{K_{\rm c}}
= L^{y_t}(a_1+c L^{y_1}+\cdots )\, ,
\end{equation}
where $a_1$ is the leading amplitude. Only terms of first order in 
$(K-K_{\rm c})$ in Eq.~(\ref{drq}) survive. From Eq.~(\ref{dQ}) one finds
that, at the transition point,
\begin{equation}
 \frac{\ln({\rm d}Q/{\rm d}K)}{\ln L} =
y_t+\frac{ \ln a_1 + (c/a_1) L^{y_1} +\cdots}{\ln L}
\label{ldQ}
\end{equation}
so that, since $y_1<0$, one expects that a plot of
$\ln({\rm d}Q/{\rm d}K)/\ln L$ versus $1/\ln L$ will yield a straight 
line for large $L$ with  an intercept $y_t$ on the vertical axis.
The data for ${\rm d}Q/{\rm d}K$, as obtained from fits to the $Q$
versus $K$ simulation results and by numerical differentiation, are 
shown accordingly in Fig.~\ref{q3rg}.

\subsection {Various results in the first-order range}
We wish to verify the results in the preceding subsection, which
indicate that the ordering phase transitions of three-state Potts models
with $z\gae 80$ are first-order.
\subsubsection{ Time evolution and histogram}
To display the discontinuous character of the transition in the model
with $z=100$ equivalent neighbors, we have recorded the behavior of the
energy of an $L=200$ system as a function of Monte Carlo time.
The system appears to display a sort of flip-flop behavior between two
states with different energies, at random intervals typically in the
order of $10^5$ Wolff cluster steps. But the fluctuations of the
higher-energy state are  still considerable which suggest that we 
should also bring the aspect of system size into consideration.

\begin{figure}
\begin{center}
\leavevmode
\epsfxsize 10.0cm
\epsfbox{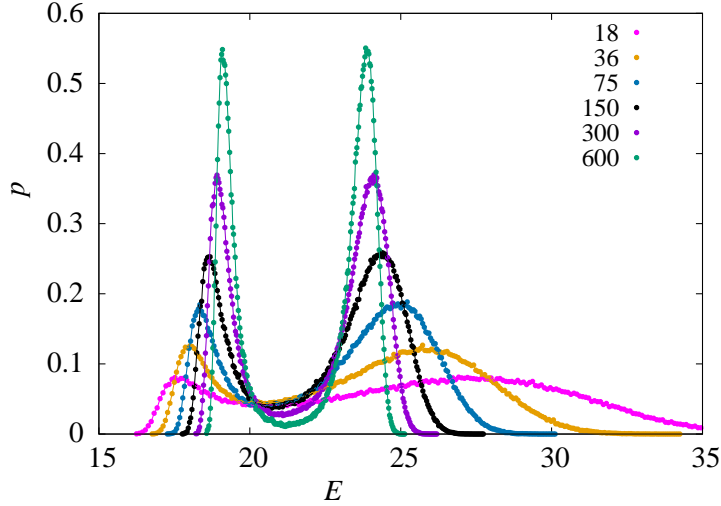}
\end{center}
\caption{Histograms of the energy distributions $p$ of finite $q=3$ Potts
models with $z=100$ equivalent neighbors. The values of the finite
sizes $L$ are shown in the figure. The couplings are chosen 
such that the two peaks are equally high. These data represent
$5 \times 10^5$ samples separated by $L/4$ single-cluster steps per
system size, except for $L=600$ where the latter number is $L/3$.}
\label{histo}
\end{figure}

Histograms of the energy are shown in Fig.~\ref{histo} for several
system sizes, taken at couplings chosen such that both maxima have the 
same height. Minor reweighting was applied to this purpose. These 
results show that the peaks become narrower and the minima between them
deeper when the system size increases. This is in accordance with
first-order behavior \cite{LK}.
\subsubsection{Hysteresis loops}
\begin{figure}
\begin{center}
\leavevmode
\epsfxsize 8.0cm
\epsfbox{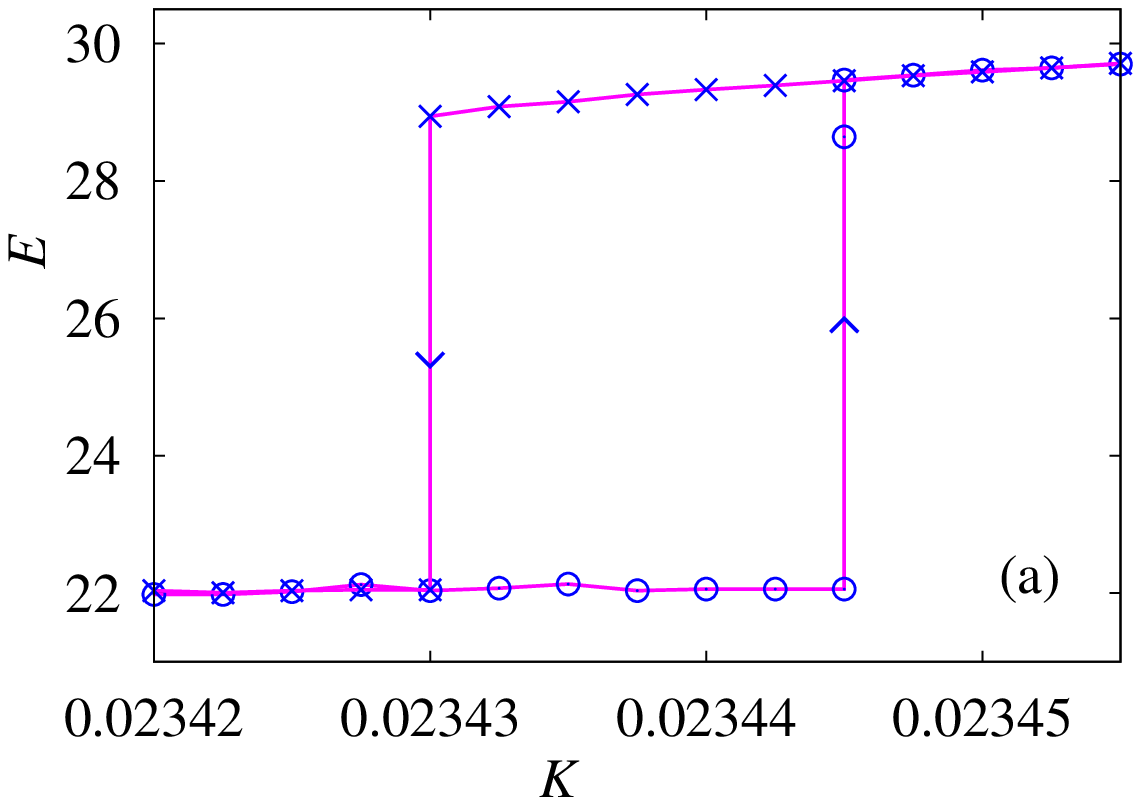}
\epsfxsize 8.0cm
\epsfbox{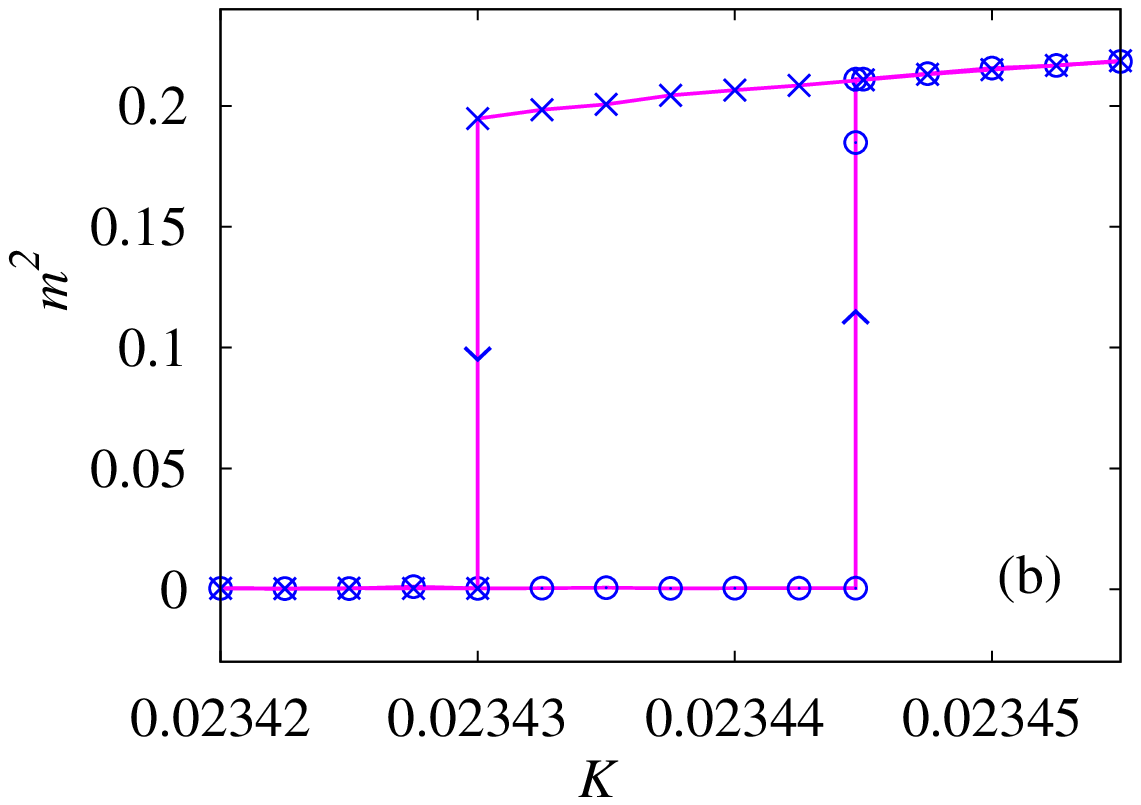}
\end{center}
\caption{Hysteresis loops of the energy (a) and the
squared magnetization (b) of the $q=3$
Potts model with 120 equivalent neighbors for system size $L=600$.
Data points are separated by $2 \times 10^5$ single-cluster steps.
A data point at the end of the observed metastability could be
obtained from intermediate results taken at smaller intervals.}
\label{em120n}
\end{figure}

We have recorded the behavior of the energy and the magnetization of the
model of an $L=600$ system with $z=120$ equivalent neighbors, while the
coupling was stepped up or down in small intervals.
The results for the energy and the magnetization are
displayed in Figs.~\ref{em120n}. The energy-like quantity $E$ is defined 
as the reduced Hamiltonian (\ref{model}) divided by $-L^2K$.
These data display clear hysteresis loops. The first-order transition
is located near $K_{\rm c}\approx 0.0234$; this is rather close to the
mean-field prediction \cite{MS,Wu} $K_{\rm c}=0.02310 \ldots$ for $z=120$.
The ranges of overlap of the two branches in Figs.~\ref{em120n} are
narrow, roughly $10^{-5}$ in $K$. While this is much smaller than the 
range of metastability according to the mean-field prediction for $q=3$,
it is naturally dependent on the system size and the simulation length
per data point.

\subsubsection{Dynamic behavior}
\begin{figure}
\begin{center}
\leavevmode
\epsfxsize 8.4cm
\epsfbox{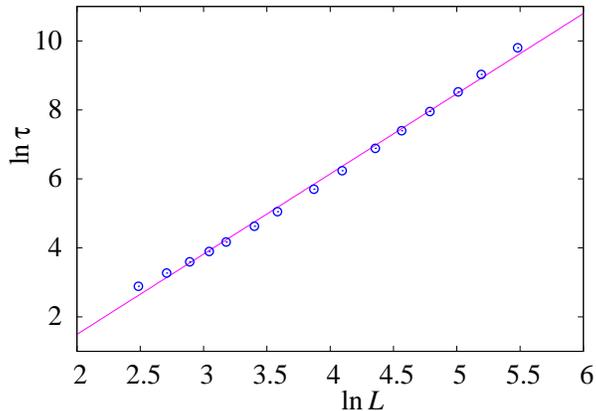}
\end{center}
\caption{Dynamic properties of the cluster simulation of the $q=3$ model
with $z=100$ equivalent
neighbors, in terms of the autocorrelation time $\tau$ versus the system 
size $L$. The use of logarithmic scales leads to a straight line with
slope ${z_d}$ if $\tau \propto L^{z_d}$. The upward curvature of the data 
is in agreement with a weakly first-order transition. The slope of the
straight line corresponds to with a dynamic exponent $z_d=2.3258$. The line
is shown for visual aid only.}
\label{ztau}
\end{figure}

Figure \ref{ztau} displays the dynamic behavior of the $q=3$ model
with $100$ equivalent neighbors at the phase transition point,
under single-cluster steps. The figure
shows the autocorrelation time $\tau$ versus the system size $L$.
The autocorrelation time unit is chosen as the number of Wolff-type
single-cluster steps equal to the inverse single-cluster size.
In the case of a critical point, one expects $\tau \propto L^{z_d}$.
The use of logarithmic scales would then lead to a straight line with
slope ${z_d}$ if $\tau \propto L^{z_d}$ in Fig.~\ref{ztau}.
The upward curvature of the line
indicates that the average cluster size does not scale algebraically
with $L$, confirming the weakly first-order character of the transition.

\section {Results for four-state Potts models}
\label{sec4}
\subsection {Auxiliary results }
\label{sec4aux}
As for the three-state model, one may attempt to determine the universal ratio
$Q_0$ from simulations of the nearest-neighbor Potts model at the exactly
known critical point. However, the logarithmic corrections for $q=4$ lead to
anomalously slow finite-size convergence and inhibit accurate numerical
analysis. Instead, we chose 
the Baxter-Wu model \cite{BW}, a model of Ising spins on the triangular
lattice, with three-spin interactions $K s_i s_j s_k$ in each triangle.
It is solved exactly \cite{BW} and belongs to the 4-state Potts
universality class, but without logarithmic corrections. 
In view of its triangular geometry, caution is needed to obtain the
universal result for $Q_0$ for models defined on a square periodic box
with the proper boundary conditions.

The invariance of boundary conditions under renormalization indicates that
value of $Q_0$ is universal, but still depending on the type of boundary
conditions. In the case of periodic boundary conditions, the periodic
images may, for instance, form a square or a triangular lattice. 
It is thus not surprising that the value of $Q_0$ was found to be
different in these two cases \cite{KBQ}; see also a confirmation by
Selke \cite{WS} and a discussion by Dohm \cite{VD}.
Furthermore, in the case that the periodic
images form a rectangular pattern, $Q_0$ is a universal function of the
aspect ratio \cite{KBQ}. In the case of a model with anisotropic couplings,
this universal function can be used to determine the equivalent geometric
anisotropy ratio \cite{KBQ,BD}.

In order to account for these effects, we chose the following numerical
approach.  We simulated Baxter-Wu systems of
$L_x \times L_y$ spins, with $L_x$ a multiple of 3, and $L_y$ a
multiple of 2, and $L_y/L_x\approx 2/\sqrt 3$. The $x$ direction is 
parallel to one set of edges of the lattice. The proper positioning
of the periodic box with respect to its periodic images was guaranteed
by choosing a square lattice representation of the triangular lattice,
with diagonal bonds added in the $(1,1)$ direction in the elementary
faces labeled with even $y$, and in the $(-1,1)$ direction in the
faces labeled with odd $y$. We employed the Wolff-like variant of an
algorithm \cite{NE} that freezes one of three sublattices, and grows a
single Ising cluster on the remaining two sublattices.
Simulations, performed at the critical point $K_{\rm c}=\half \ln(1+\sqrt 2)$,
involved 45 system sizes with $3\leq L_x\leq 240$, with a number of
samples in the order of $10^9$ for $L \leq 72$ and $10^8$ for $L > 72$. 
The periodic boxes defined above are rectangular,
with aspect ratios that are only approximately equal to 1. Therefore
the aspect ratio was included in the fit formula Eq.~\ref{drq} for the
finite-size data as follows:
\begin{equation}
Q(L_x,L_y)=Q_0+b_1 L^{-1}+b_2 L^{-7/4}+b_3 L^{-2}+c_1 a^2(L_x,L_y)+ \ldots
\label{QBW}
\end{equation}
where $L\equiv \sqrt{L_x A_y}$, with $A_y\equiv\sqrt{3/4}L_y$ the actual
size of the rectangular periodic box in the $y$ direction. The aspect
ratio is parametrized by $a\equiv (L_x-A_y)/\sqrt{L_x^2+A_y^2}$.
The correction exponent $y_1=-1$ was strongly suggested by the finite-size
data, and is equal to the difference between the first two magnetic
exponents $y_{h2}-y_{h}$ \cite{BN}. The exponent $y_2=-7/4$ is equal to
$2-2y_h$ and may arise from the analytic part of the susceptibility.
Also the term with $y_3=-2$ helped to reduce the fit residuals,
enabling satisfactory fits for $Q$ down to a minimum system size of $L_x=6$.

We included an independent determination of $Q_0$ from simulations of
the dilute $q=4$ Potts model on the square lattice at the estimated
fixed point  $K_f=1.45790$, $D_f=2.478438$, which is very close to
the value reported in Ref.~\onlinecite{QDB}. We simulated $L \times L$ 
systems for a number of finite sizes $L=4,$ 5, ...., 80.
Since logarithmic corrections are absent at the fixed point, we used
Eq.~(\ref{QBW}) to fit to the finite-size data, but without the
term accounting for the aspect ratio. The fits behave very similar
to those for the Baxter-Wu model, and the results for $Q_0$ of both
models agreed satisfactorily.

We performed several other fits, by including a correction with an
exponent $y_2=-2.5$ instead of $-1.75$, and with various subsets of
fixed correction exponents. After discarding the fits with a too large
residual $\chi^2$, the results are consistent with our final estimate
$Q_0=0.81505 \, (15)$ where the error estimate is twice the statistical
margin of the average of the preferred fits for the two models.
This value of $Q_0$ will be helpful in the analysis of the results for
the $q=4$ equivalent-neighbor models.

In addition to the universal ratio $Q_0$, we also investigate the 
universal ratio $c_1/b$ mentioned in Sec.~\ref{sec2}, by means of
simulations of a modified Baxter-Wu model. The model remains
self-dual when the couplings $K_{\rm up}$ and $K_{\rm down}$ in the
up- and down triangles are made different. The self-dual line is
located at $\sinh 2K_{\rm up}\sinh 2K_{\rm down}=1$.
For $K_{\rm up} \ne K_{\rm down}$ the model shifts away from the $q=4$
fixed point, and thus acquires logarithmic corrections \cite{genbw}.
The direction of its shift is away from the nearest-neighbor model, 
into the first-order range. Since the ratio $K_{\rm up}/K_{\rm down}$
can be chosen arbitrarily, we can arrange it such that for our range of
$L$ values the finite-size-scaling behavior of the model is determined
by the renormalization flow in the vicinity of the fixed point.
Thus the value of the marginal field $v$ in Eq.~(\ref{fq4}),
as well as that of the parameter $c$ in Eq.~(\ref{Qq4c}), remains small. 
Then, we may assume that higher-order terms with $c_2$, $c_3$, etc.~ in
that equation may be neglected. Under this assumption we attempt to
determine the universal ratio $c_1/b$ from a fit of Eq.~(\ref{Qq4c})
to the Monte Carlo results for $Q$, taken at the self-dual point for
a suitable value $K_{\rm up}/K_{\rm down}$.
We simulated the model with $K_{\rm up}/K_{\rm down}=2$ at the 
self-dual point, for 27 system sizes $6 \leq L \leq 120$. Most of
the simulations were rather short, with a few times $10^7$ samples,
but we also included long runs with about $10^9$ samples for $L=24$,
48 and 72. Good statistics is necessary, because the differences of the
finite-size data for $Q$ and those of the fixed point, which we wish to
analyze, are still quite small.
Satisfactory least-squares fits could be obtained on the basis of 
Eq.~(\ref{Qq4c}) for system sizes down to $L_x=6$. We obtain
$c_1=-0.0049~(2)$ and $b=0.102~(6)$, from which we estimate the
universal ratio $c_1/b=-0.048$.

\subsection {Critical points}
\label{sec4kc}
We estimated the critical points and the temperature exponent $y_t$,
as well as $Q_0$, from the Monte Carlo data for the Binder ratio
for several values of $z$ in the $q=4$ medium-range Potts model. 
As a preliminary analysis, we fitted Eq.~(\ref{drq}) to the finite-size
data for $Q$, with the values of $Q_0$ and $y_t$ left free. The
correction exponents were fixed at $y_1=-1$ and $y_2=-7/4 $. The fit
results are shown in Table \ref{pot4}. While these results are
inaccurate as a measure of the universal quantities, they provide 
information how the nature of the phase transition depends on $z$. 
For $z \lae 12$, the estimates of $y_t$ are smaller than the exact value
$y_t=3/2$,
as is usually the case for $q=4$ Potts-like models with short-range
interactions \cite{RS,BN82,ATTTI}. The estimates of the Binder ratio
are clearly too large in comparison with the universal value
$Q_0=0.81505 \, (15)$ as listed in Sec.~\ref{sec4aux}. These discrepancies
are explained by logarithmic factors, such as in Eq.~(\ref{Qq4c}), which
are not taken into account in these fits. These differences decrease when
$z$ increases, signaling a decrease of the marginal field $v$. The results
for $z \gae 20$  indicate that the model resides in the first-order range.
This is probably also the case for $z=20$, since the $y_t$ estimates
exceed 3/2, with an increasing trend for large $L$, suggesting
crossover to the discontinuity fixed-point value $y_t=2$.
Since the fixed-point value of $z$ seems to lie between 12 and 20, we
have included a model with $z=16$ equivalent neighbors. We realized this
by including only four of the eight neighbors at a distance $R=\sqrt 5$,
with coordinates $(x,y)=(2,1),(-1,2),(-2,-1),(1,-2)$. This preserves the
fourfold rotational symmetry of the local interacting environment.
\begin{table}[h]
\centering
\caption{\label{pot4} Binder ratio $Q$ and thermal exponent
$y_t$ as estimated from simulations of the medium-range $q=4$
Potts model. These results suggest that the tricritical point between
the critical and first-order range occurs between $z=12$ and 20.
The error margins, quoted as 2 times the standard
deviation of the statistical analysis, are not realistic because
logarithmic correction factors are omitted in this analysis.
Moreover, the errors for $z=60$ may be underestimated because of
slow dynamics in the first-order range.  The third column shows the
number of millions of samples taken for each data point
as specified by $K,L$. A number of $K$ values near $K_{\rm c}$ was
chosen for each $L$, typically varying between 6 for $L<20$ and 1 for
the largest values of $L$.
}
\vskip 5mm
\begin{tabular}{|c|c|c|l|l|l|l|l|}
\hline
$z$& $L$  & Ms  &$K_{\rm c}$ &  $Q$      &$y_t$      & $y_1$     & $y_2$ \\
\hline
 4 &12-240&  8 &1.09862   (1)&0.840   (2)& 1.418  (5)&$-$1       &$-$7/4 \\
 8 &12-224&  8 &0.49098   (2)&0.836   (2)& 1.431  (4)&$-$1       &$-$7/4 \\
12 &12-224& 30 &0.30625   (2)&0.828   (3)& 1.490 (10)&$-$1       &$-$7/4 \\
16 &12-224& 30 &0.222856  (2)&0.814   (1)& 1.529 (10)&$-$1       &$-$7/4 \\
20 &12-224& 10 &0.175842  (2)&0.805   (1)& 1.610 (10)&$-$1       &$-$7/4 \\
24 & 8-120& 12 &0.144523  (2)&0.795   (1)& 1.70   (4)&$-$2       &$-$3   \\
28 & 8-96 & 12 &0.122812  (2)&0.788   (1)& 1.82   (6)&$-$2       &$-$3   \\
36 & 8-84 & 15 &0.094528  (2)&0.780   (1)& 1.91   (5)&$-$2       &$-$3   \\
44 & 8-48 & 25 &0.076826  (4)&0.781   (6)& 2.02   (5)&$-$2       &$-$3   \\
60 &12-44 & 20 &0.055921  (2)&0.804   (8)& 2.04   (5)&$-$2       &$-$3   \\
\hline
\end{tabular}
\end{table}

We have also determined the first derivative of $Q$ with respect to
the coupling $K$ at criticality, similarly as for $q=3$. In the
critical range one thus expects, in principle, convergence of
$(\ln({\rm d}Q/{\rm d}K)/\ln L$ to $y_t=3/2$, but the presence of a
marginal field leads to corrections behaving as an inverse logarithm
of $L$, so that the available range of system sizes is insufficient
for an accurate result.
\begin{figure}
\begin{center}
\leavevmode
\epsfxsize 15.0cm
\epsfbox{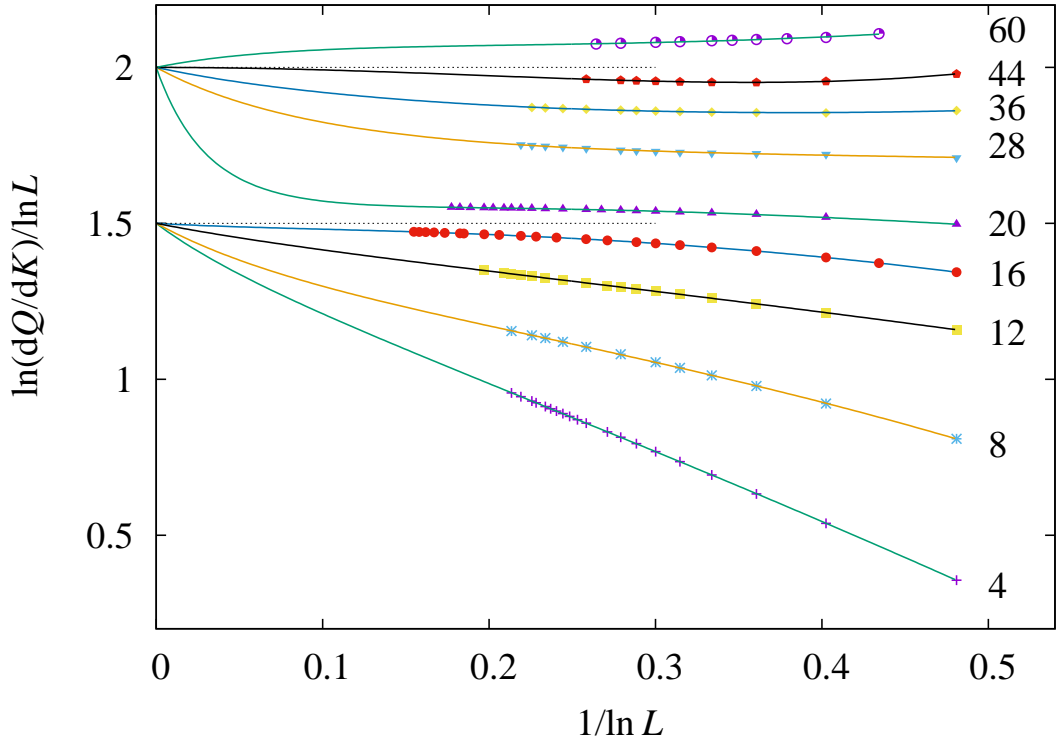}
\end{center}
\caption{Finite-size dependence of the derivative at $K_{\rm c}$ of the
Binder ratio $Q$ of the 4-state Potts model with respect to $K$.
The data points for each value of $z$ are connected by by lines
which are intended for visual aid, and for display of possible
extrapolations to $L=\infty$. The values of $z$ are shown in the figure
for each curve.  This way of plotting should lead, for critical models
with sufficiently large $L$, to linear behavior and convergence to the 
temperature exponent $y_t=3/2$. However, due to the presence of
logarithmic corrections, such behavior may only be observed in practice
when the marginal field vanishes. The data in this figure suggest that
this is the case near $z=16$. }
\label{dQ4}
\end{figure}
Nevertheless, the data for $(\ln({\rm d}Q/{\rm d}K)/\ln L$
versus $1/\ln L$, shown in Fig.~\ref{dQ4} are sufficiently clear to
demonstrate that the $z=16$ model lies close to the $q=4$ fixed point, 
and signals the boundary  between the short-range
behavior for $z < 16$ and first-order behavior for $z > 16$.

In an attempt to obtain more accurate estimates of the critical points,
$Q_0$ and $y_t$ were fixed at their expected values, and a fit formula
based on Eq.~(\ref{drq4}) was used for $z \leq 20$. The $z=20$ model still
seems to be rather close to the fixed point. However, except for $L=16$, 
it appears that the accuracy of the fits is limited, because the
higher-order logarithmic terms do not converge satisfactorily. As a
result, the parameters $b$ and $c_1$, purportedly describing the
marginal scaling field, are poorly determined. Instead, we define,
on the basis of Eqs.~(\ref{Qq4c}) and (\ref{drq4}), a measure of the
marginal field as the sum of the logarithmic terms at $K_{\rm c}$
\begin{equation}
\Delta Q(L) \equiv \sum_k c_k/(1-b\ln L)^k=Q(K_{\rm c},L)-
\sum_j b_j L^{y_j} -Q_0 \, .
\label{delq}
\end{equation}
At a constant finite size $L$, this quantity represents the scaling 
function $Q(u)$ as a function of the distance $u$ to the fixed point.
Unlike the individual amplitudes, the sum of the logarithmic terms is
well determined by the fit, at least within the range of $L$ covered
by the data. We chose $L=80$ where the power-law corrections, and their
error margins, are relatively small. 
The results for $\Delta Q(80) $ are included in  Table \ref{pot42}.
For $z \geq 24$ we used a fit formula with a different value $Q_0=0.8$
and without logarithmic terms. Power-law corrections are included 
with exponents as shown in Table \ref{pot42}.
The distance to the discontinuity fixed point is
purportedly approximated by $\Delta Q(L) \equiv Q(K_{\rm c},L) -Q_0$ 
at a sufficiently large size $L=80$. \\
\begin{table}[htbp]
\centering
\caption{\label{pot42} Critical points $K_{\rm c}$ for $q=4$ as derived
from fits with the Binder ratio $Q_0$ and thermal exponent $y_t$ fixed as
shown in the table.  Error margins are quoted as twice the standard
deviation of the statistical analysis.
The exponents of the power-law corrections were fixed at the values shown
in the table, except for $y_1$ in the range $24 \leq z \leq 44$ where
this exponent was left free in the fit.
}
\vskip 5mm
\begin{tabular}{|r|r|l|l|l|l|l|l|}
\hline
$z$ &$L_{min}$&$K_{\rm c}$&$Q_0$ &$y_t$ & $y_1$     & $y_2$&$\Delta Q(80)$ \\
\hline
 4  & 6  &ln\,3 (exact) & 0.81505  &3/2   &$-$1       &$-$7/4  &~~0.0346   \\
 8  &12  &0.49097    (1)& 0.81505  &3/2   &$-$1       &$-$7/4  &~~0.0245   \\
12  & 8  &0.306252   (1)& 0.81505  &3/2   &$-$1       &$-$7/4  &~~0.0136   \\
16  &12  &0.222856   (1)& 0.81505  &3/2   &$-$1       &$-$7/4  &$-0.0010 $ \\
20  &12  &0.175843   (1)& 0.81505  &3/2   &$-$1       &$-$7/4  &$-0.0072 $ \\
24  & 8  &0.144523   (1)& 4/5      &2     &$-$0.3  (1)&$-$2    &$-$0.005   \\
28  & 8  &0.122812   (1)& 4/5      &2     &$-$0.6  (1)&$-$2    &$-$0.010   \\
36  & 8  &0.094531   (2)& 4/5      &2     &$-$0.6  (1)&$-$2    &$-$0.018   \\
44  & 8  &0.076829   (2)& 4/5      &2     &$-$0.8  (1)&$-$2    &$-$0.012   \\
60  &12  &0.055919   (2)& 4/5      &2     &$-$1       &$-$2    &$-$0.004   \\
\hline
\end{tabular}
\end{table}
The accuracy of the values of $\Delta Q(80)$, shown in the last column
of Table \ref{pot42}, is estimated as about 0.001. The results in the
range $4 \leq z \leq 20$ clearly display a change of sign of the
marginal field near $z=16$. The results in the range $24 \leq z \leq 60$
indicate that the finite-size scaling function of $Q$ describing the 
crossover from the merged fixed point to the discontinuity fixed point
goes through an extremum before approaching the limit $Q=4/5$.
\subsection {Hysteresis loop}
\begin{figure}
\begin{center}
\leavevmode
\epsfxsize 8.0cm
\epsfbox{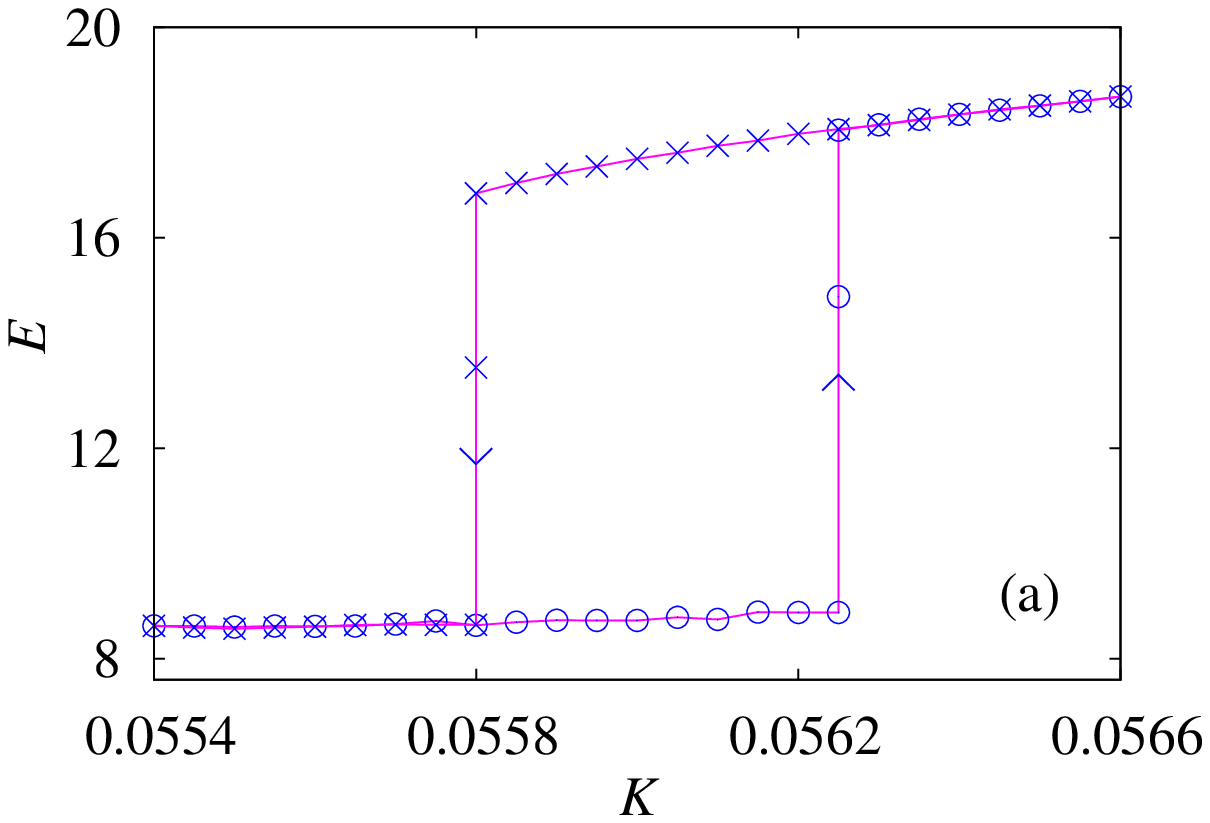}
\epsfxsize 8.0cm
\epsfbox{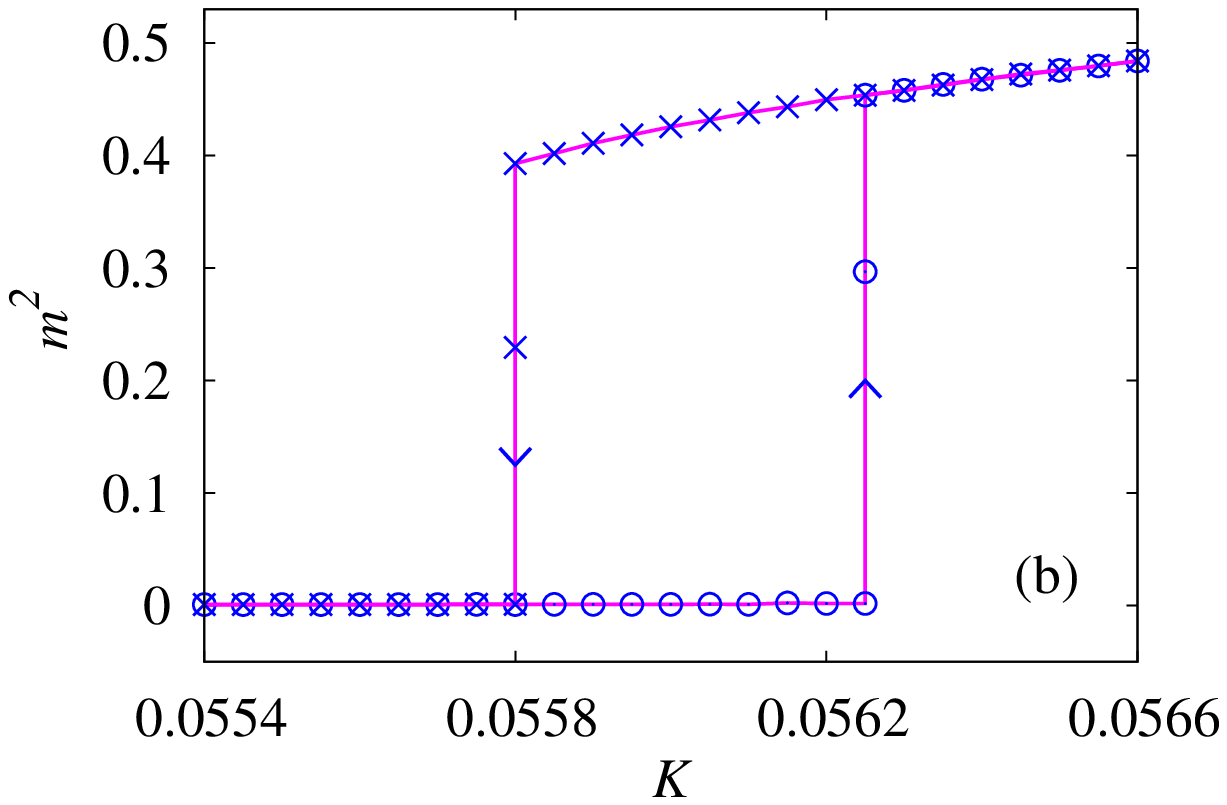}
\end{center}
\caption{Hysteresis loops of the energy (a) and
the squared magnetization (b) of the $q=4$ model
with 60 equivalent neighbors and finite size $L=120$.
The energy-like quantity $E$ is equal to the expectation value of
the reduced Hamiltonian (\ref{model}) divided by $-L^2K$.
Data points are separated by $2 \times 10^5$ single-cluster steps.
Two data points at the end of the observed metastability could be
obtained from intermediate results taken at smaller intervals.}
\label{q460nem}
\end{figure}

For $q=4$ we have  determined the behavior of the energy and the
magnetization of an $L=120$ system with $z=60$ equivalent neighbors,
while the coupling was stepped up or down in small intervals.
We find very clear hysteresis loops, which are displayed in
Figs.~\ref{q460nem}.
The first-order transition takes place near $K_{\rm c}=0.0559$, not far 
from the mean-field prediction $K_{\rm c}=0.0593$ for the $q=4$ model
with $z=60$ interacting neighbors.

\section {Discussion and conclusion}
\label{sec5}
\begin{figure}
\begin{center}
\leavevmode
\epsfxsize 7.9cm
\epsfbox{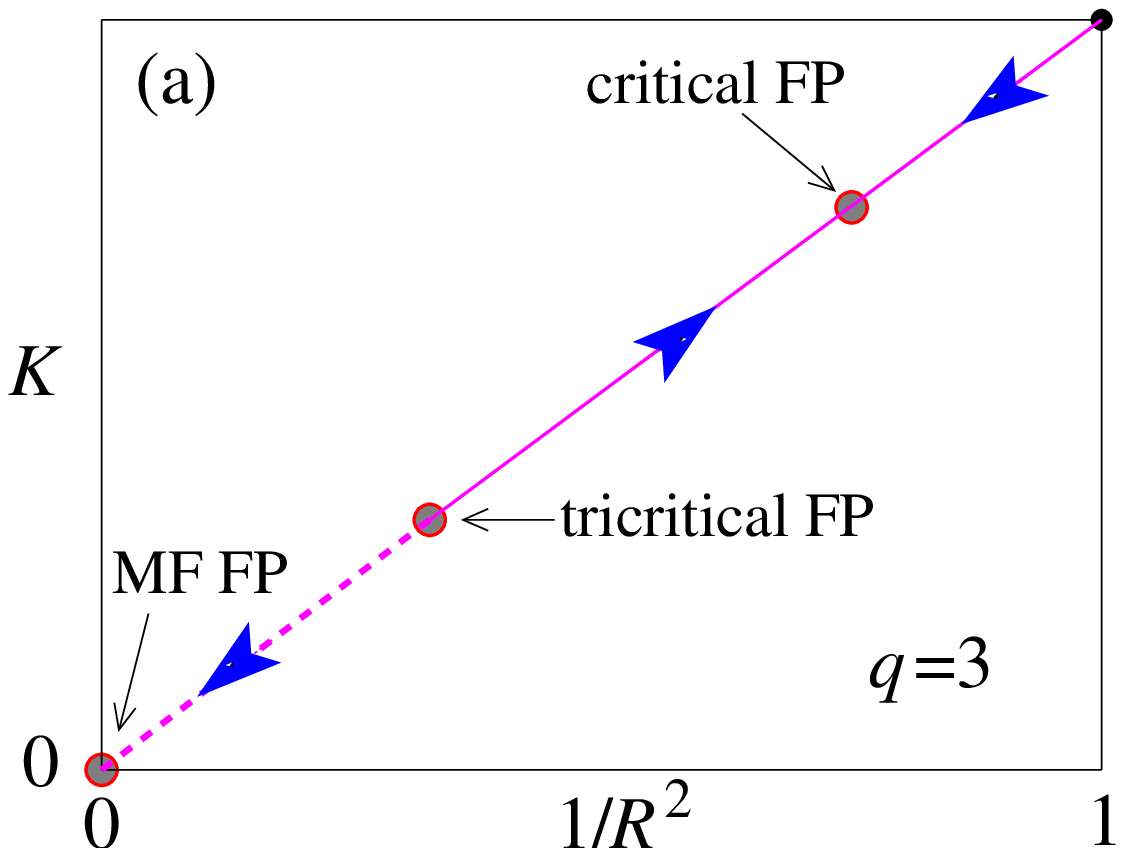}
\epsfxsize 7.5cm
\epsfbox{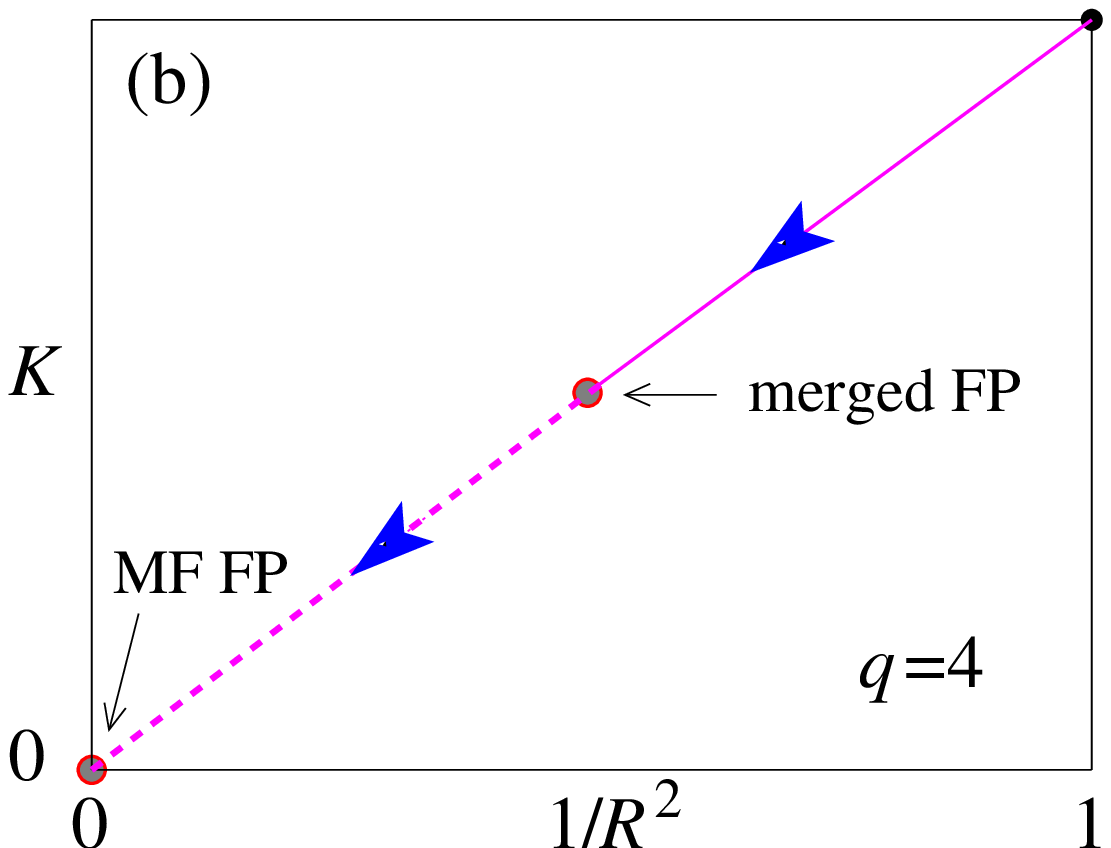}
\end{center}
\caption{Renormalization flow of the three-state (a) and
of the four-state (b) Potts model along the line of
ordering transitions.}
\label{rgq34}
\end{figure}

The results in Sec.~\ref{sec3} indicate that, for the $q=3$ Potts model,
the renormalization flow is as shown in Fig.~\ref{rgq34} (left-hand side),
i.e., the role of the interaction range is similar to that of the
fugacity of the vacancies in the dilute Potts model \cite{NBRS}.
Our results indicate that the $q=3$ Potts model with $z=80$ lies close
to the tricritical fixed point in Fig.~\ref{rgq34}, and that the critical
fixed point corresponds with a value of $z$ between 8 and 12.
Also the results for the $q=4$ model in Sec.~\ref{sec4} display this
analogy: increasing the range of the interactions has a similar effect
as dilution in the nearest-neighbor $q=4$ Potts model \cite{NBRS,QDB}.
 
Thus our results are well described by the renormalization flow diagram
that follows when the critical and tricritical fixed points in
Fig.~\ref{rgq34} merge into a single fixed point that is marginally
irrelevant on the short-range side and marginally relevant on the
long-range side \cite{NBRS}, as shown in Fig.~\ref{rgq34}. Since the
marginal field is absent in the Baxter-Wu model \cite{BW}, that model
faithfully reproduces the expected scaling behavior a the merged fixed
point. Our results in Sec.~\ref{sec4kc} indicate that also the $q=4$
Potts model with $z=16$ lies close to the merged fixed point in
Fig.~\ref{rgq34}.

These results for the $q=3$ and 4 Potts model stand in a strong contrast
with the Ising case $q=2$, where the effects of vacancies and of an
increased interaction range are  different.  The Ising tricritical point
such as present in the dilute Ising model is absent in the $q=2$ model with 
medium-range interactions \cite{LBB}, as illustrated in Fig.~\ref{ising}.
There is only a gradual crossover, with mean-field behavior only in the
limit $R \to \infty$. Ising universality applies for all finite $R$. 

An obvious question not answered in the present work is how the present 
work can be generalized to non-integral values of $q$, i.e., the
medium-range random-cluster model \cite{KF}. Self-consistent solution
in the mean-field limit $z \to \infty$ shows that, for $q<2$, the
critical behavior of this model is the same as that of the mean-field 
percolation model, with critical exponents $\beta=1$, $\gamma=1$ and
$\delta=2$.  For this range of $q$, we conjecture that the mean-field
fixed point is unstable with respect to finite values of $z$.
We thus expect that, for $q<2$, the universal behavior is that of the
short-range $q$-state random-cluster model, for all finite ranges $R$
of interaction.

\acknowledgments
We acknowledge useful discussions with A.~D. Sokal about the value
of $Q_0$ in the first-order range. H.~B. is grateful for the hospitality 
extended to him by the Faculty of Physics of the Beijing Normal University.
This research was supported by the National Natural Science Foundation
of China under Grant No.~11275185 (YD), No.~ 11175018 (YL,WG), and by the
Fundamental Research Funds for the Central Universities of the Ministry
of Education (China).
% under Grant No.~2340000034.\\


\begin{thebibliography}{99}
\bibitem{Yuhai}
The value $Q_0=q/(q+1)$ follows if, at coexistence, the $q$ magnetized
phases are equally probable as the disordered phase. This value is
confirmed by an exact asymptotic $L \to \infty$ analysis of finite
Potts models in the mean-field limit $R \to \infty$.
\end{thebibliography}
\end{document}